# Sources Variability With Planck LFI

L.Terenzi[1], M.Bersanelli[2], C.Burigana[1], R.C.Butler[1], G.De Zotti[3], N.Mandolesi[1], D.Mennella[4], G.Morgante[1], M.Sandri[1], L.Valenziano[1], F.Villa[1]

[1]*Istituto Te.S.R.E./ CNR – Bologna – Italy,*
[2]*Università di Milano – Milano – Italy,*
[3]*Osservatorio Astronomico di Padova – Padova – Italy*
[4]*Istituto di Fisica Cosmica/ CNR – Italy*
*On behalf of the Planck Collaboration*

**Abstract.** Planck LFI (Low Frequency Instrument) will produce a complete survey of the sky at millimeter wavelenghts. Data stream analysis will provide the possibility to reveal unexpected millimeter sources and to study their flux evolution in time at different frequencies. We describe here the main implications and discuss data analysis methods. Planck sensitivities typical for this kind of detection are taken into account. We present also preliminary results of our simulation activity.

## INTRODUCTION

The LFI (Low Frequency Instrument) [1] on board of the Planck satellite is designed and optimized to measure primary anisotropies in the CMB (Cosmic Microwave Background); the images that it will produce at 30, 44, 70 and 100 GHz will have an unprecedented combination of sky coverage, calibration accuracy, freedom from systematic errors, stability and sensitivity. The LFI data will represent also a good opportunity to study the astrophysics of extragalactic radiosources. In particular, LFI is expected to be efficient in the detection of extragalactic radiosources in active phases characterized by high emission levels (BL Lac, blazar). We collected informations on typical flux intensities, spectral variabilities and light curves of these objects (see, e.g., [2]) to properly exploit the experiment observation strategy, based upon a full sky coverage in periods of about six months at four frequency channels.

## PLANCK LFI SENSITIVITY

Starting from the LFI sensitivities at different frequencies (Planck Low Frequency Instrument, Instrument Science Verification Review, October 1999, private reference), considering the main properties of the Planck scanning strategy [3] and the LFI beam positions on the telescope field of view [4], we are able to evaluate the averaged instrumental sensitivities of the LFI array for the study of variable sources on different time-scales and the number of relevant observations with the quoted sensitivities (Table 1; see [5] for further details). The global rms noise in the Planck data streams is the sum in quadrature of all the relevant contributions, assumed independent. The CMB, Galaxy and extragalactic source confusion noises per beam ($FWHM^2$) vary respectively from about 250, 100 and 60mJy to 190, 7 and 20mJy, when the frequency goes from 30 to 100 GHz.

Therefore, in the final recovery of radiosource flux variability we can take advantage from the knowledge of diffuse component fluctuations at the highest and middle HFI frequency channels (at high resolution) since the CMB dominates over the astrophysical confusion noise. The global noise, relevant in this context, is then dominated by the LFI data stream receiver noise.

The Planck sensitivity depends also on the sky position, mainly on the ecliptic colatitude $\theta_e$; the baseline scanning strategy implies a sensitivity, expressed in terms of sensitivity averaged on the sky, approximated by the law $(\sin\theta_e/\sin50°)^{1/2}$; sources at high ecliptic latitudes are observed much longer than those at low latitudes.

| TABLE 1. Instrumental Performances And Number Of Measurement For Typical Time-Scales. | | | | |
|---|---|---|---|---|
| Period | >14 Months (Aux. Data) | 14 Months | 1-6 days/sin50° | Few-12 hours/sin50° |
| $\sigma_{noise}$ (mJy) at 30 GHz | 13.4 | 19.0 | -- | -- |
| $N_{meas}$ at 30 GHz | 1 | 2 (3) | -- | -- |
| $\sigma_{noise}$ (mJy) at 44 GHz | 20.5 | 29.0 | 35.7 - 50.0 | -- |
| $N_{meas}$ at 44 GHz | 1 | 2 (3) | 2 | -- |
| $\sigma_{noise}$ (mJy) at 70 GHz | 28.0 | 39.6 | -- | 68.6 |
| $N_{meas}$ at 70 GHz | 1 | 2 (3) | -- | 3 |
| $\sigma_{noise}$ (mJy) at 100 GHz | 32.2 | 45.5 | 71.0 – 132.9 | 108.5 – 132.9 |
| $N_{meas}$ at 100 GHz | 1 | 2 (3) | 6 - 7 | 2 - 3 |

## SIMULATIONS

We have simulated the LFI observations of a number of sources located at different positions in the sky, in order to evaluate the impact of different source observation durations.

We have considered a representative set of beam positions among those recently simulated for the current baseline [4] corresponding to the LFI feed horns 27 (at 30 GHz), 25 (at 44 GHz), 21 (at 70 GHz), and to the feed horns 2, 6 and 9 (at 100GHz).

Time ordered data (TODs), expressed in term of simulated antenna temperature, have been generated (see Fig. 1) for sources with different fluxes (1, 3 and 5 Jy). The relationship between the source flux and the observed antenna temperature is used to translate the simulated antenna temperature TODs into source flux evaluations.

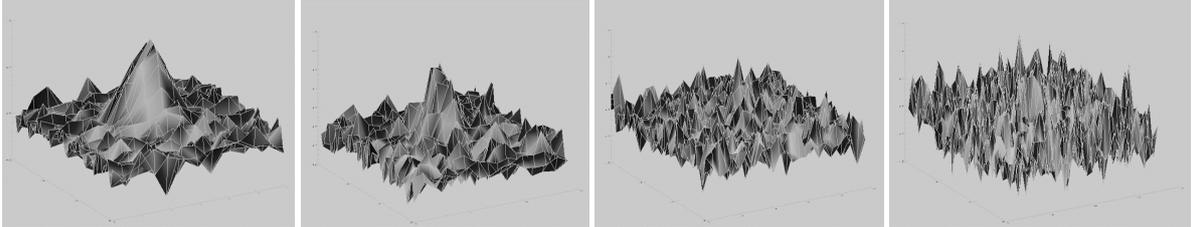

**FIGURE 1.** TODs from simulation of transit of a source of 3 Jy at medium ecliptic latitude in feed horns respectively at 30, 44, 70 and 100 GHz.

Next, we analyze the instrument efficiency for a preliminary source flux reconstruction. As evident, the 100 GHz receivers are too noisy to allow a satisfactory flux reconstruction by using a single beam, so in subsequent analyses we have considered the combination of six beams (12 receivers).

Of course, starting from TODs, this calculation brings out to ugly flux fluctuations when the beam center points to regions relatively far from the sources. Therefore, in order to efficiently reconstruct the source flux we have to optimize the choice of the region extent around the source in a way more or less strict according to the source luminosity and the channel sensitivity.

We firstly assume a good knowlegde of the beam pattern and neglect possible pointing errors.

Estimates of the optimized number of samples to properly reconstruct the source flux at low and high ecliptic latitudes are shown in Tables 2 and 3 for fluxes of 1 and 5 Jy. Along the scan circle direction, we find that it is advantageous to work with three samples around the source; in that direction the sampling interval is equivalent to 1/3 of the FWHM, so we find advantageous to use an angular scale of about one FWHM. $\Delta\phi_{eff}$ reported in the table is the optimal interval considered along the spin axis re-pointing direction multiplied by $\sin\theta_e$; it ranges between one and two times the beam FWHM. The accuracy in the flux recovery is also reported in the tables.

We have further analyzed the effect of the pointing uncertainty upon flux reconstruction, assuming as a reference a pointing error of 1arcmin at $1\sigma$ level. The impact on source flux reconstruction is shown in columns 2-5 in the table 4, where, in analogy with previous panels, we report relative errors in flux at low and high ecliptic latitudes (LEL and HEL). The present error estimates are based on 30 simulations.

The pointing error implies also a systematic error in the main beam in-flight recovery during the mission. Given the current estimates on the relative uncertainty in beam resolution evaluation (ΔFWHM/FWHM of about 0.007 at 30 GHz and 0.022 at 100 GHz [6,7]) introduced by such a level of pointing uncertainty, we find a relative error, ΔF/F, on source flux recovery from a TOD sample of about 0.007 (0.028) for samples at $\theta \cong 1\sigma_{beam}$ ($2\sigma_{beam}$) from the source for the 30 GHz channel, and 0.022 (0.089) at $1\sigma_{beam}$ ($2\sigma_{beam}$) for the 100 GHz channel. Note as, in general, these effects induced by pointing uncertainty are relevant for the highest frequency channels because of their better resolution and become comparable to the noise sensitivity in the case of bright sources. Therefore, a pointing accuracy significantly better than $\cong$ 1arcmin at 1σ level, at least by a factor 2 as in the current LFI requirements [6], is extremely important not only for the LFI cosmological aim but also for accurate radiosource variability studies.

**TABLE 2. Sampling Width and Accuracy in Flux Recovery of a Source at Low Ecliptic Latitude.**

| Frequency (GHz) | $\frac{\Delta F}{F}$ (1 Jy) | $\frac{\Delta\phi_{eff}}{FWHM}$ | $\frac{\Delta F}{F}$ (5 Jy) | $\frac{\Delta\phi_{eff}}{FWHM}$ |
|---|---|---|---|---|
| 30 | 0.085 | 0.7 | 0.016 | 0.6 - 1.5 |
| 44 | 0.120 | 1.8 | 0.026 | 1.7 |
| 70 | 0.239 | 1.6 | 0.056 | 1.6 |
| 100 | 0.41 | 1 | 0.092 | 1 |
| 100 (6 beams) | 0.131 | 1.2 | 0.033 | 2 |

**TABLE 3. Sampling Width and Accuracy in Flux Recovery of a Source at high Ecliptic Latitude.**

| Frequency (GHz) | $\frac{\Delta F}{F}$ (1 Jy) | $\frac{\Delta\phi_{eff}}{FWHM}$ | $\frac{\Delta F}{F}$ (5 Jy) | $\frac{\Delta\phi_{eff}}{FWHM}$ |
|---|---|---|---|---|
| 30 | 0.051 | 1.5 - 1.8 | 0.011 | 1.6 |
| 44 | 0.083 | 1 | 0.017 | 0.9 – 1.3 |
| 70 | 0.193 | 1.1 | 0.035 | 1.2 |
| 100 | 0.243 | 1 | 0.058 | 1 |
| 100 (6 beams) | 0.090 | 0.8 | 0.021 | 1.8 |

**TABLE 4. Effect of Pointing uncertainty on the Accuracy in Flux Recovery.**

| Frequency (GHz) | $\frac{\Delta F}{F}$ Rms (LEL) | $\frac{\Delta F}{F}$ Half Disp. (LEL) | $\frac{\Delta F}{F}$ Rms (HEL) | $\frac{\Delta F}{F}$ Half Disp. (HEL) |
|---|---|---|---|---|
| 30 | 0.0082 | 0.016 | 0.0046 | 0.0087 |
| 44 | 0.0146 | 0.0237 | 0.0095 | 0.0161 |
| 70 | 0.0246 | 0.0501 | 0.0198 | 0.0419 |
| 100 | 0.050 | 0.1073 | 0.0267 | 0.0537 |
| 100 (6 beams) | 0.0176 | 0.0417 | 0.0117 | 0.02407 |